\documentclass[final,10pt]{svjour2}
\usepackage{graphicx}
\usepackage{rotating}
\usepackage{amssymb}
\usepackage{mathptmx}
\usepackage[numbers]{natbib}
\makeatletter
\journalname{Journal of Low Temperature Physics}

\bibpunct{}{}{,}{s}{}{,}

\renewcommand{\vec}[1]{{\mbox{\boldmath$#1$}}} 

\newcommand{\mat}[1]{\ensuremath{\mathbf #1}}   

\newcommand{\be}{\begin{equation}}
\newcommand{\ee}{\end{equation}}
\newcommand{\ba}{\begin{eqnarray}}
\newcommand{\ea}{\end{eqnarray}}

\begin{document}

\newcommand{\hdblarrow}{H\makebox[0.9ex][l]{$\downdownarrows$}-}

\title{Optimization and analysis of code-division multiplexed TES microcalorimeters}

\author{
J.~W.~Fowler$^1$,
W.~B.~Doriese$^1$,
G.~Hilton$^1$,
K.~Irwin$^1$,
D.~Schmidt$^1$,
G.~Stiehl$^1$,
D.~Swetz$^1$,
J.~N.~Ullom$^1$,
and
L.~Vale$^1$}

\institute{1:Quantum Sensors Project, National Institute of Standards and Technology, 325
  Broadway MS 817.03, Boulder, CO 80305, USA \\
Tel: +1 303--497--3990\\
\email{joe.fowler@nist.gov} \\
\emph{Official contribution of the National Institute of Standards and Technology; not
subject to copyright in the United States.}
}

\date{22.07.2011}

\maketitle


\begin{abstract}

  We are developing code-division multiplexing (CDM) systems for TES arrays with the
  goal of reaching multiplexing factors in the hundreds.  We report on x-ray
  measurements made with a four-channel prototype CDM system that employs a
  flux-summing architecture, emphasizing data-analysis issues.  We describe an
  empirical method to determine the demodulation matrix that minimizes
  cross-talk.  
%
%
  This CDM system achieves energy resolutions of between 2.3\,eV and 3.0\,eV FWHM at
  5.9\,keV\@.

\end{abstract}

\section{Introduction}
Transition-edge sensors\cite{irwin95} (TESs) are used in many different applications
to sense photons over an enormous energy range.
They are the leading detector technology in millimeter-wave and submillimeter
astronomy, where focal planes of thousands of sensors have been
fielded\cite{swetz11,carlstrom11} and show great promise for astronomical x-ray
spectroscopy.  In the laboratory, TES devices have achieved the best energy
resolution of any non-dispersive device for single photons in the gamma and x-ray
bands---better than one part in one thousand.

One critical technology for building arrays of thousands of detectors
has been cryogenic signal multiplexing, which reduces the number of electrical
connections between stages at different temperatures by allowing multiple
detectors to share wiring.  As future projects grow to larger device
counts and employ faster detectors, they will require new multiplexing techniques to
minimize both readout noise and instrument complexity.

\begin{figure}
\begin{center}
\parbox[b]{\linewidth}{
  \includegraphics[%
    width=.67\linewidth,
    keepaspectratio]{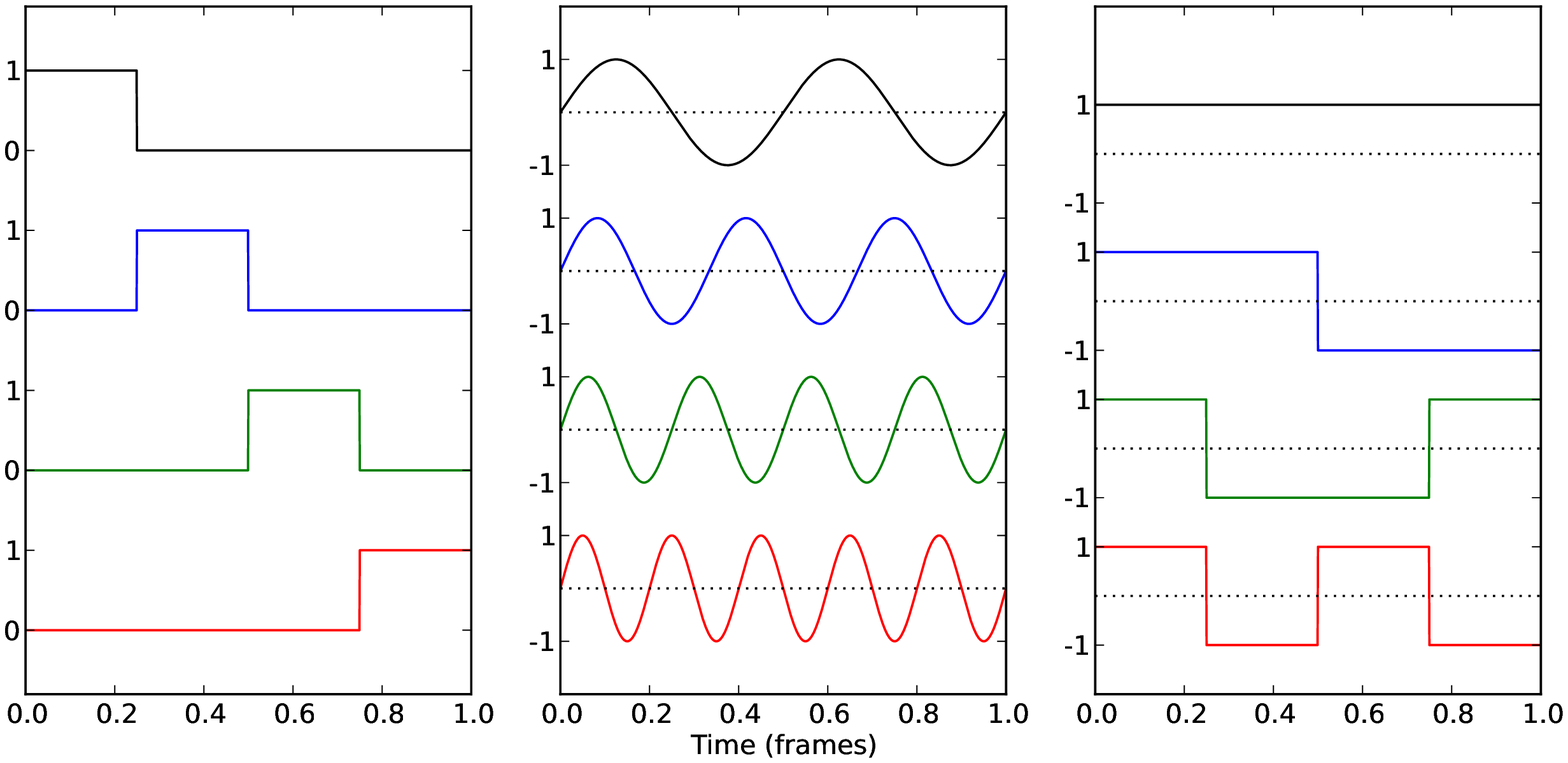}
  \hfill
  \includegraphics[%
    width=0.27\linewidth,
    keepaspectratio]{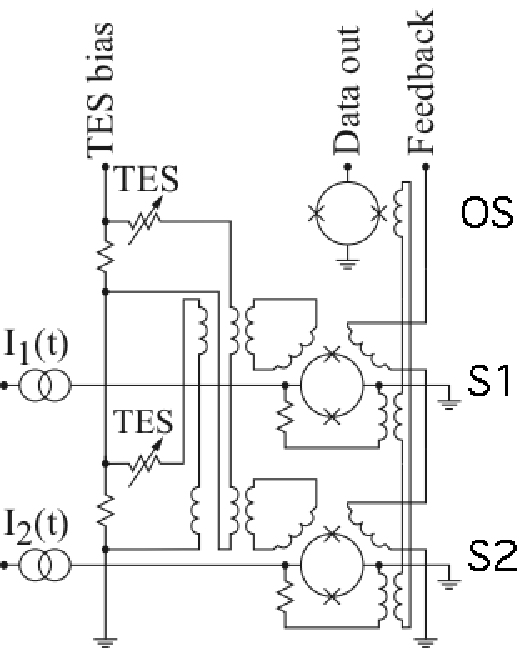}
}
\end{center}
\caption{(Color online) \emph{Left:} Three sets of orthogonal basis functions for
  multiplexing four pixels: time-division (boxcars), frequency-division (sinusoids),
  and code-division (Walsh codes).  In each multiplexing scheme, the four detector
  outputs are multiplied by one of the four given modulation functions.
  \emph{Right:} A two-channel code-division multiplexer that uses flux summation.  Labels
  ``OS'', ``S1'', and ``S2'' refer to the output SQUID and SQUID switches 1 and 2.
  The magnetic flux in SQUID switch 1 is proportional to the sum of the two TES
  currents, while the flux in switch 2 is proportional to their difference (indicated
  by the oppositely-oriented inductors).  When current is applied at either $I_1(t)$
  or $I_2(t)$, the output SQUID couples to either the sum or the difference of TES
  currents.  We report on this CDM design, generalized to four TES detectors
  and an equal number of SQUID switches. }
\label{fig:basis_functions}
\label{fig:cdm_schematic}
\end{figure}

Two multiplexing schemes currently used with TES arrays are time-division
multiplexing (TDM)\cite{chervenak} and frequency-division multiplexing
(FDM).\cite{cunningham}  In this work, we present the first x-ray measurements with
TES microcalorimeters read out with code-division multiplexing (CDM). This
multiplexing approach combines many advantages of the other two, as argued in earlier
works.\cite{karasik01,podt2002} Figure~\ref{fig:basis_functions} (left panel)
compares the modulation functions used in TDM, FDM, and CDM for a four-channel
system.  In TDM, one detector is coupled to the output at a time.  In CDM, all
detectors are coupled at all times---with time-varying signs---removing the
``multiplex disadvantage'' of TDM, in which the aliased SQUID noise increases as the
square root of $N$, the number of multiplexed detectors.  CDM therefore has the
potential to permit higher device speed, or to increase the number of devices
multiplexed on one line, or both.  We describe the design and characterization of a
four-channel prototype and demonstrate that the energy resolution achieved while
measuring 5.9\,keV x-rays matches the resolution achieved through TDM readout.

\section{A 4-channel CDM demonstration system}


We have measured x-rays with microcalorimeters optimized for the 1 to 10\,keV energy
range, read out through a demonstration chip fabricated at NIST containing a
four-channel CDM circuit.
%
%
The code-division multiplexer operates by the principle of flux
summation.\cite{irwin2010} The right panel of Figure~\ref{fig:cdm_schematic} shows a
simplified two-channel model of the multiplexer.  As in a TDM system, $N$ detectors
couple their current into $N$ switching SQUIDs, whose outputs are summed by a single
output SQUID\@.  In a TDM system, the detectors and switching SQUIDs are paired so
that each detector couples into only one switching SQUID; in the CDM system, all $N$
detectors couple into all $N$ SQUIDs, each detector having a different pattern of
positive and negative couplings.  For an appropriate set of $N^2$ coupling
polarities, the detectors' patterns are mutually orthogonal over the period required
to read all $N$ switching SQUIDs.  The Walsh functions\cite{walsh23} provide one such
basis.

The TDM and flux-summed CDM systems are used identically; in both, the $N$ SQUID
switches are activated singly in rapid succession.  We can therefore use CDM chips as
drop-in replacements for the TDM, with identical warm electronics to multiplex the
signals.  This interchangeability has enabled both rapid testing of the CDM
prototype and direct comparisons of CDM with TDM results.

\begin{figure}
\begin{center}
\includegraphics[%
  width=0.75\linewidth,
  keepaspectratio]{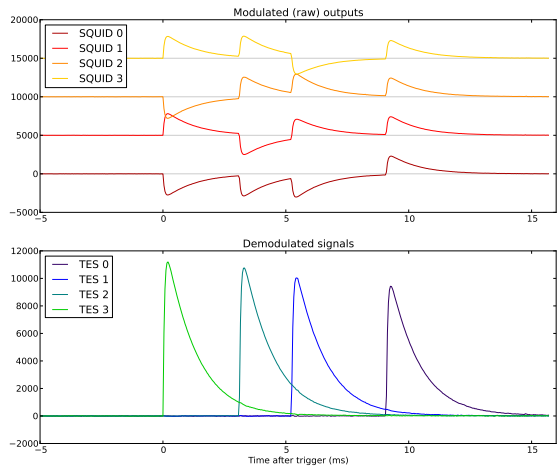}
\end{center}
\caption{(Color online) Sample data from observing manganese fluorescence light for
  21\,ms.  The raw data from the four SQUID switches (\emph{top}) and the same data
  demodulated to recover the current from each of the four TES detectors
  (\emph{bottom}) are shown.  The vertical scale is arbitrary; the upper panel
  offsets each curve for clarity.  This unusual time interval was selected to
  demonstrate all four modulations in one figure; only a small fraction of pulse
  records contain even two photons coinciding in such a short window.}
\label{fig:raw_cdm}
\end{figure}

\section{Characterization of the CDM data}

We used CDM to read out a TES array measuring x-rays emitted by a manganese target
bombarded with wide-band brehmsstrahlung radiation.  Manganese fluoresces with a pair
of K$\alpha$ lines\cite{holzer} at 5.9\,keV separated by 11.1\,eV.  We estimate the
energy resolution of the calorimeters from the observed line shapes.

Figure~\ref{fig:raw_cdm} shows a brief excerpt of the data from the 4-channel CDM
system.  The top panel shows the raw data as recorded by the system, one record from
each of the four switching SQUIDs.  They have been modulated by the Walsh functions
designed into the CDM chip.  Inversion of the Walsh coding yields the separate TES
currents---as shown in the bottom panel of Figure~\ref{fig:raw_cdm}.


\begin{figure}
\begin{center}
\includegraphics[%
  width=.8\linewidth,
  keepaspectratio]{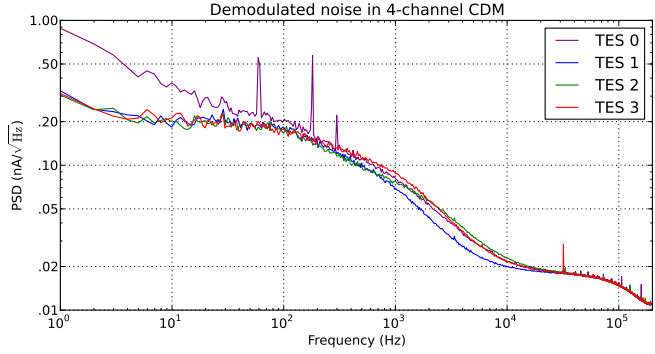}
\end{center}
\caption{(Color online) The square root of the noise power spectral density in
  demodulated four-channel CDM data, averaged over 90 seconds with no x-ray sources.
  TES~0--the one unmodulated detector---has the highest noise level below 200\,Hz,
  owing to gain drifts in second and third stage SQUIDs and pickup of the 60\,Hz AC
  power line frequency (and at two odd harmonics).  Modulation by Walsh functions
  removes most of this extra noise from the signal band in the other detector
  channels.  The in-frame linear time correction described in
  Section~\ref{sec:crosstalk} suppresses noise above 100\,kHz in all channels.}
\label{fig:noise_psd}
\end{figure}

One advantage of modulation by Walsh functions is that all but one detectors' signals
are differenced at multiples of the modulation frequency.  Upon demodulation,
low-frequency drifts and noise at the AC power line frequency and its overtones are
moved out of the signal band, provided their source is upstream of the modulator
(such as in the warm electronics).  This effect can be seen in
Figure~\ref{fig:noise_psd}, where only TES~0 (the unmodulated device) has excess
noise below 100\,Hz.

\subsection{Solving for the demodulation matrix}

One challenge in CDM is computation of the true demodulation matrix \mat{D} required
to find the current through each of the $N$ detectors given the $N$ multiplexed
outputs.  We define $D_{ij}$ as the contribution of the current from TES $i$ to
output channel $j$ .  By design, this matrix consists entirely of $+1$ and $-1$
elements:
\be \mat{D} = \left[
\begin{array}{rrrr}
 1 &  1 &  1 &  1 \\
-1 &  1 &  1 & -1 \\
-1 & -1 &  1 &  1 \\
-1 &  1 & -1 &  1 \\
\end{array}
\right]
\label{eq:walsh}
\ee
In flux-summed CDM, the modulation coefficients are wired into the design by the
lithographic traces in the multiplexer and cannot be adjusted after fabrication.
Because the exact couplings depend on details of the SQUID switches and the geometry
of the coils that convert current into magnetic flux, \mat{D} differs somewhat from
its design value.  We estimate the matrix empirically and have found that it is
consistent over many weeks of operation.  When needed, it can be computed from any
sufficiently long calibration data set.

\begin{figure}
\begin{center}
\includegraphics[ width=0.9\linewidth, keepaspectratio]{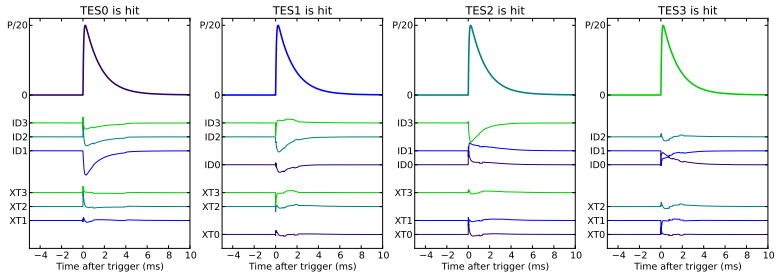}
\end{center}
\caption{(Color online) Reduction in cross-talk through refining the
  demodulation matrix.  Each panel shows the pulses caused by the TES cited at the
  top.  The highest curve shows the pulse shape for the primary signal, scaled down
  by a factor of 20 for better comparison to cross-talk.  Below this, three curves
  labeled ``ID'' show the unwanted cross-talk response in the \emph{other} three
  channels, assuming an integer-only demodulation matrix. The lowest three curves
  show the cross-talk (``XT'') when the empirically derived demodulation is applied
  instead.  }
\label{fig:crosstalk}
\end{figure}

We refine our estimate of \mat{D}, starting from the integer-only matrix
(Equation~\ref{eq:walsh}).  The data required for the refinement are the mean current
signals $s_{ij}(t)$ from demodulated channel $i$ when device $j$ (and no other) is
hit with a photon, plus the noise power spectrum $n_j$ of each device.
Figure~\ref{fig:crosstalk} shows an example of all sixteen signals with the $i=j$
cases plotted in the top of each panel and the twelve cross-talk ($i\ne j$) cases
below.  The sampled functions of time are each converted to single values by
the usual process of optimal filtering.  A correction matrix \mat{C} is built from
the filtered values: $C_{ij}$ is the response in channel $i$ to a photon hitting
detector $j$.  The correction \mat{C} will be close to the identity if \mat{D} was a
good initial estimate.  The revised demodulation matrix is then $\mat{D'} =
\mat{C}^{-1}\ \mat{D}$.  In practice, we find that elements of $\mat{D'}$ are
generally within 1\% of $\pm 1$.  We have not found it necessary to repeat this
procedure iteratively.  The three lowest signals in each panel of
Figure~\ref{fig:crosstalk} show the mean cross-talk in the demodulated channels after
applying the refined $\mat{D'}$ instead of the integer-only $\mat{D}$.  This
correction reduces the cross-talk between channels from the 1\% level to only a few times
$10^{-4}$.

\subsection{Other sources of cross-talk}
\label{sec:crosstalk}


Even properly estimated, $\mat{D}$ can remove cross-talk only if it is proportional
to the primary pulse.  Some forms of cross-talk, including a time-offset effect and
magnetic coupling, are proportional to the time derivative of a pulse.
If all cross-talk can be reduced to a level much lower than the TES energy
resolution, then an array can operate at high photon rates with no degradation
due to cross-talk.

The time-offset effect stems from the fact that the $N$ modulated signals (upper
panel of Figure~\ref{fig:raw_cdm}) are measured not simultaneously but in succession
over some 1 to 10\,$\mu$s.  This is not ideal when one of the input signals
changes rapidly, as at the beginning of a photon pulse. 
The time offset effect can be greatly reduced by linearly interpolating all modulated
data streams between their actual sample times to estimate their values at the reference
time, which is the same for all channels.
 We apply
this ``in-frame linear time correction'' to all data shown in this report before
applying (or estimating) the demodulation $\mat{D}$, markedly reducing the size of
the remaining cross-talk during the $\sim 100\,\mu$s rising edge of each pulse.
Although the linear assumption can be generalized to higher orders through the approach
of Savitzky-Golay filtering\cite{numrec}, we find quadratic time correction not to
be useful in the current system. The most important departure from linearity is the corner
in the signal at the pulse onset, which higher-order polynomials do not model well.


Even with the corrections described so far, cross-talk in a form proportional to the
time derivative of a pulse remains a concern.  It probably results from magnetic
coupling between channels and is just strong enough to limit the energy resolution at
high photon rates.  Research continues on this cross-talk mechanism and on how to
mitigate it in hardware and in later analysis.

\subsection{Performance of the multiplexed microcalorimeters}

\begin{figure}
\begin{center}
\includegraphics[%
  width=1\linewidth,
  keepaspectratio]{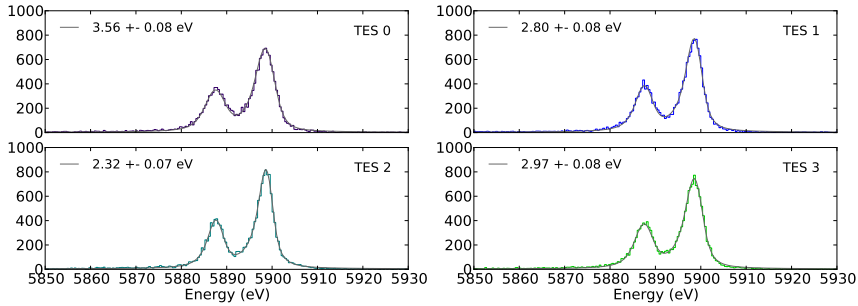}
\end{center}
\caption{(Color online) The measured energy distributions of Mn K$\alpha$
  fluorescence (histograms) and the best fit distributions (smooth curves) assuming
  Gaussian-distributed resolutions of the given full width at half-maximum.  The
  TESs are read out through a 4-channel flux-summing CDM.  Vertical units are
  photons per 0.4\,eV-wide bin.  TES~0 is the unmodulated device and is
  expected not to perform as well as the others.}
\label{fig:spectra}
\end{figure}

The proof of detector and multiplexer performance is the energy resolution achieved
in real measurements; we employed the manganese fluorescence lines at 5.9\,keV to
estimate our detector resolutions.  Typical observations involved one to ten photons
per second per TES detector and lasted up to three hours.  After making the
corrections described above, we processed triggered data records roughly 15\,ms long
with optimal filtering\cite{optfiltering} and a simple drift-correction proportional
to variations in the baseline (pre-triggered) signal level of each TES.

We have found that the simplest method of fitting measured energy distribution to the
expected line shape---by minimizing Pearson's $\chi^2$---produces results biased
towards lower (better) energy resolution.  We therefore always perform a full
maximum-likelihood fit to estimate energy resolutions, taking account of the
Poisson-distributed nature of counts in a histogram.\cite{histogram-fit} For the data
shown here with $\sim 10^4$ photons per histogram, use of the maximum-likelihood fit
makes the FWHM energy resolution approximately 0.05\,eV worse than the $\chi^2$
biased fit does.  A larger bias is found when fewer photons are measured.

Figure~\ref{fig:spectra} shows the best energy resolution accomplished over a long
observation.  The raw photon rate of 3.6\,Hz per detector was maintained for 130
minutes.  Each histogram contains between 17,000 and 18,000 photons in the
Mn~K$\alpha$ energy range depicted, after all data-quality cuts.  As expected, the
three modulated detectors have better energy resolutions than TES~0.  Their mean
resolution of 2.70\,eV exactly matches the mean resolution we have achieved through
TDM with eight identical detectors in the same cryostat.
For these spectra, we have vetoed records in which more than one of the four
detectors are hit, as cross-talk remains just large enough to have an impact on
resolution.  This cross-talk cut improves the energy resolutions by 0.05\,eV on
average while removing some 15\% of the records.  In experiments with higher photon
rates, the improvement in resolution with the cut is larger, but the price paid in
lost data is also larger.  Further reduction of cross-talk is vital to future
high-rate experiments.

\section{Future directions}

The CDM demonstration chip used in the present experiments also contains an
eight-channel multiplexer, which we will test in the near future.  In addition, a
32-channel device has been fully designed.  Reaching higher multiplexing factors will
require improved techniques for minimizing cross-talk between calorimeters.

A further development that promises to allow in-focal-plane multiplexing and a
greatly reduced wire count is the current-summed CDM.\cite{irwin11} In that concept,
the TES current polarity is modulated by SQUID-based superconducting double-pole
double-throw switches.  This approach allows the CDM basis functions to be
stored in firmware rather than being patterned onto the SQUID chip.  It is also
compatible with microwave multiplexing.

We plan to use CDM as a simple upgrade for TDM readout systems, with significantly
better performance at high multiplexing factors.  These results show that the
demonstration device with our current analysis techniques achieves energy resolutions
as good as those found in similar detectors operated through TDM\@.



\newcommand{\apl}[1]{\emph{Appl.\ Phys.\ Lett.\ }{\bf #1}}
\newcommand{\ltd}{\emph{Proc.\ LTD-14}}

\end{document}